\documentclass[a4paper,preprint,showkeys]{amsproc}
\usepackage{mathrsfs}
\usepackage[utf8]{inputenc}
\usepackage{mathtools}
\usepackage{amsmath,booktabs}
\usepackage[T4, OT1]{fontenc}
\usepackage[arrow, matrix, curve]{xy}
\usepackage{newunicodechar}
\usepackage{graphicx}
\usepackage{bm}
\usepackage[dvips]{epsfig}
\usepackage{rotating}
\usepackage{amsmath,amssymb,amsfonts,amsthm}
\usepackage{amsmath,amscd}
\usepackage{subfig}
\usepackage{caption}
\usepackage{pst-all}
\usepackage{dcolumn}
\usepackage[thinc]{esdiff}
\usepackage{amsthm}
\usepackage{dsfont}
\usepackage{amssymb}
\usepackage[hyphens]{url} 
\usepackage[dvips]{graphicx}
\usepackage{xparse}
\usepackage{physics}
\usepackage{mathrsfs}
\makeatletter
\@namedef{subjclassname@2024}{\textup{2024} Mathematics Subject Classification}
\makeatother
\theoremstyle{plain}

\theoremstyle{definition}

\theoremstyle{remark}
 
 \numberwithin{equation}{section}

\newcommand{\eg}{e.\,g.\ }
\newcommand{\ie}{i.\,e.\ }

\title[quantized magnetic flux]{Magnetic flux as a quantized Lorentz pseudoscalar\\ and its relation to electric charge quantization.}
\author[Belardinelli]{Cyril Belardinelli\\Kantonsschule Solothurn KSSO\\Switzerland} 
\address{ 
Kantonsschule Solothurn\\
Solothurn\\
Switzerland}
\email{cyril.belardinelli@ksso.ch}
\begin{document}
\vspace{18mm} \setcounter{page}{1} \thispagestyle{empty}
\begin{abstract}
In this paper, we re-examine the well-known question of why electric charges exist only in quantized portions. In this context, we revisit the motion of a charged particle in a field-free region around a current-carrying solenoid. Solving the corresponding Schrödinger equation leads to a simultaneous quantization condition for the magnetic flux and the electric charge. We also demonstrate the Lorentz invariance of this condition by showing that the magnetic flux behaves like a pseudoscalar under Lorentz transformations.
\end{abstract}
\maketitle
\section{\label{history}Introduction}
The question of why electric charge occurs in quantized portions \textit{is one of the greatest mysteries in physics}\cite{Jackson:1998}. Dirac \cite{dirac:1931} demonstrated that the existence of magnetic charges leads to the quantization of electric charge. However, to this day, magnetic monopoles have not been detected. Of course, one could simply accept the quantization of charge as a given fact of Nature without asking for deeper reasons. Such reasons may not even exist. 
But from a theorist's perspective, this is not an appealing approach.\\
When Maxwell passed away in 1879 he left his equations not in the highly symmetric form we know them today but rather as twenty equations in twenty variables. It was largely thanks to Oliver Heaviside (1850-1925) and independently Heinrich Hertz (1857-1894) on the Continent that the Maxwell equations were brought into their now familiar reduced form using vector operators (in Gaussian units);
\begin{eqnarray*}
&&\curl \vb{E}(\vb{x},t)+\frac{1}{c}\,\dot{\vb{B}}(\vb{x},t)=0,\qquad \,\,\, \qquad \div \vb{E}(\vb{x},t)=4\pi\, \rho(\vb{x},t) \\
\\
&&\curl \vb{B}(\vb{x},t)-\frac{1}{c}\,\dot{\vb{E}}(\vb{x},t)=\frac{4\pi}{c}\,\vb{j}(\vb{x},t),\quad \div \vb{B}(\vb{x},t)=0
\end{eqnarray*}
Interestingly, to Maxwell, the potentials (the electric potential $\bold{\varphi}$ and the magnetic vector potential $\bold{A}$), not the fields $\bold{E}, \bold{B}$, played the central role in electromagnetism \cite{maxwell1954}.
The relation between potentials and fields is given by the well-known formulae;
\begin{equation*}
\vb{E}=-\grad{\varphi}-\pdv{\vb{A}}{t},\qquad \vb{B}=\grad \cross \vb{A}
\end{equation*}
For Maxwell, the vector potential $\vb{A}$ had the meaning of a momentum. He called $\vb{A}$ the \textit{electrokinetic momentum} which to him was \textit{the} fundamental quantity in electromagnetism. 
Hertz\footnote{Hertz is known as the discoverer of electromagnetic waves, but he was a gifted theoretician as well.} and Heaviside held diametrically opposed views. Both were staunch advocates of fields; they wished to banish Maxwell’s potentials from electromagnetism. With the advent of quantum mechanics, however, potentials have regained their importance\footnote{In Feynman's words \cite{feynman2010}: \textit{In the general theory of quantum electrodynamics, one takes the vector and scalar potentials as the fundamental quantities in a set of equations that replace the Maxwell's equations: $\vb{E}$ and $\vb{B}$ are slowly disappearing from the modern expressions of physical laws; they are being replaced by $\phi$ and $\vb{A}$}.}; the Aharonov-Bohm (A-B) effect shows that a charged particle is affected by the vector potential, despite being confined to a region in which the magnetic field is zero \cite{bohm:1959}. Nevertheless, the potentials $\varphi$ and $\bold{A}$ are not directly measurable; one can only ever measure potential differences. After all, quantum mechanics is gauge-invariant, meaning that under a gauge transformation, the wave function acquires an unobservable phase. However, there is a quantity that is accessible to measurement and that is also the key quantity in the A-B-effect; namely magnetic flux.\\
This work focuses primarily on this quantity. Specifically, we investigate its behavior under Lorentz transformations showing that magnetic flux is a Lorentz Pseudoscalar. Furthermore, we argue that magnetic flux is a generally quantized quantity, just like electric charge. 
As a consequence, we point out the possibility that a quantized magnetic flux also entails the quantization of electric charge.\\
\\
In this paper, we use the relativistically covariant form of Maxwell's equations given as;
\begin{eqnarray*}
\partial_{\mu}{F}^{\mu \nu}&=&\frac{4\pi}{c}\,j^{\nu}\\
\partial_{\mu}{\tilde{F}}^{\mu \nu}&=&0
\end{eqnarray*}
where $j^{\nu}=(c\rho, \vb{j})$ denotes the electric 4-current density. 
The electromagnetic field tensor $F^{\mu \nu}$ in matrix form is given by;
\begin{equation*}
{F}^{\mu \nu}=\partial^{\mu} A^{\nu}-\partial^{\nu} A^{\mu}=\begin{pmatrix} 0& -E_x & -E_y& -E_z \\
        E_x & 0 &-B_z & B_y\\
 E_y & B_z &0 & -B_x\\
 E_z & -B_y &B_x & 0\end{pmatrix}
\end{equation*}
The dual tensor $\tilde{F}^{\mu \nu}$ in matrix form is defined as;
\begin{equation*}
\tilde{F}^{\mu \nu}=\frac{1}{2}\epsilon^{ \mu \nu \alpha \beta}{F}_{\alpha \beta}=\begin{pmatrix} 0& -B_x & -B_y& -B_z \\
        B_x & 0 &E_z & -E_y\\
 B_y & -E_z &0 & E_x\\
 B_z & E_y &-E_x & 0\end{pmatrix}
\end{equation*}
where $\epsilon^{\mu \nu \alpha \beta}$ denotes the total antisymmetric Levi-Cività tensor;
\begin{eqnarray*}
\epsilon^{\mu \nu \alpha \beta} = \begin{cases}
+1 & \text{if}\,\, ( \mu, \nu,\alpha, \beta)\,\,\text{is an even permutation of}\,\, (0, 1, 2, 3) \\
-1 & \text{if}\,\, (\mu, \nu, \alpha, \beta)\,\, \text{is an odd permutation of}\,\, (0, 1, 2, 3) \\
0 & \text{otherwise}
\end{cases}
\end{eqnarray*}
\section{\label{}Is Magnetic flux a generally quantized quantity?}
The concept of flux quantization is well known in the context of type II superconductors. In these superconductors, a magnetic field $B$ with a strength between two critical values, \ie $H_{c1}<B< H_{c2}$,  partially penetrates into the superconductor. The magnetic field is concentrated in well separated vortices of diameter $\lambda$ carrying one unit of magnetic flux;
\begin{equation*}
\Phi_{0}=\frac{hc}{2e}
\end{equation*}
It is only in this specific context that we usually speak of magnetic flux quantization, but certainly not in a general situation, \eg in free space. In what follows, we will adopt a more general perspective by means of a very simple line of reasoning. In the following, we re-examine a simple situation that can be found in many textbooks of Quantum mechanics such as \eg Griffiths \cite{griffiths1995}.\\
\\
\begin{figure}[h!]
\centering
{\includegraphics[width=0.7\linewidth]
{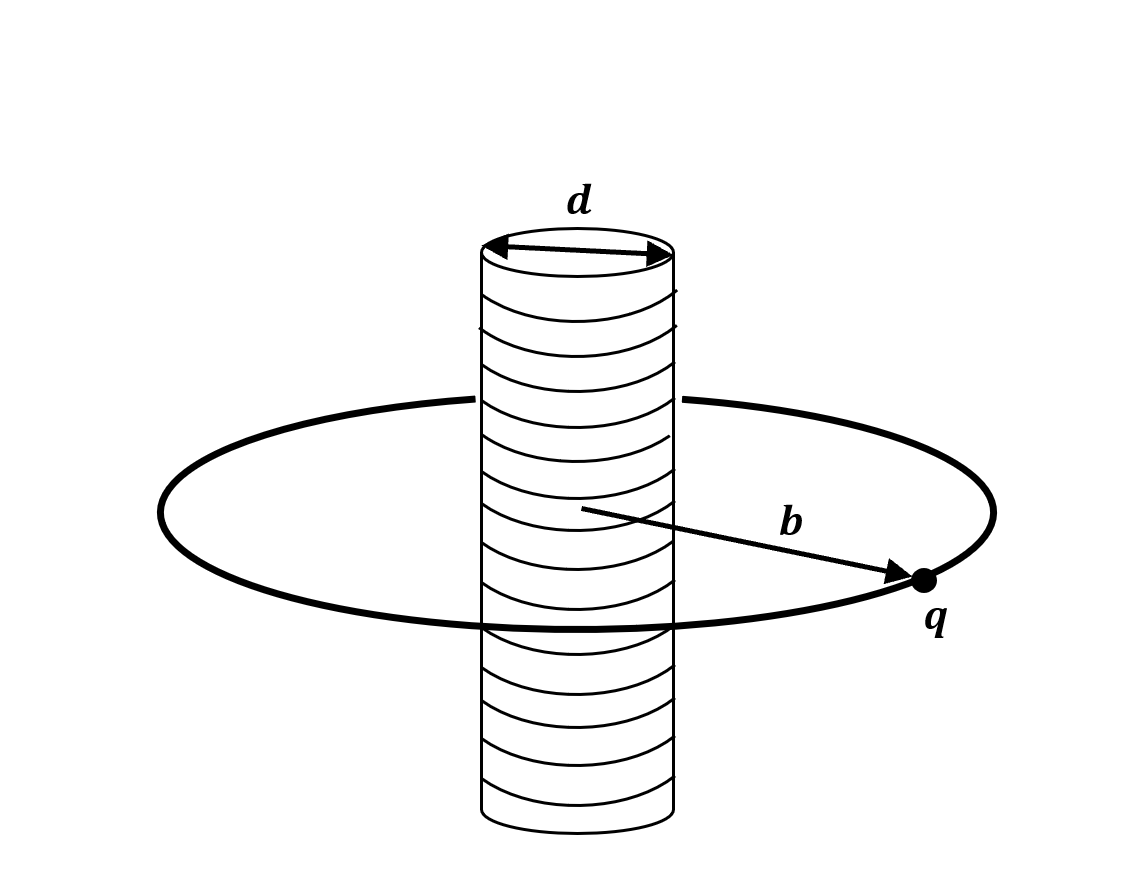}}
\caption{The electric charge orbits outside a long solenoid ($d\ll2b$) through which an electric current flows. Since the solenoid is assumed to be very thin and very long, the point charge moves in a region with negligible magnetic flux density $\vb{B}$. However, the magnetic vector potential $\vb{A}$ is still present; it cannot be eliminated by any gauge transformation.}
\label{fig1}
\end{figure}
Imagine a particle constrained to move in a circle of radius $b$. Along the axis runs a very long solenoid of diameter $d\ll2b$ carrying a magnetic field $\bold{B}$ due to the electring current passing through it. We can imagine a very long (infinitely long) solenoid so that the magnetic field outside the solenoid is arbitrary small. Remarkably, the vector potential cannot be made to vanish under any gauge transformation.\\ In the Coulomb gauge $\div{\vb{A}}=0$, the vector potential is;
\begin{eqnarray*}
\vb{A}_{\text{in}}&=&\frac{\Phi}{2\pi a^2}(-y,x,0), \qquad r<d/2\\
\vb{A}_{\text{out}}&=&\frac{\Phi}{2\pi r^2}(-y,x,0), \qquad r>d/2
\end{eqnarray*}
In theory, one can let the diameter $d$ of the solenoid approach zero while simultaneously increasing the magnetic field strength $\abs{\vb{B}}$, such that the magnetic flux through the solenoid remains unchanged. In this limiting case, a single flux line remains.\\
We will now solve the Schrödinger eigenvalue equation $\mathcal{H}\psi=E\psi$ for the charged particle on the circular ring. Here, we follow closely the textbook of Griffiths \cite{griffiths1995}, but we will draw an additional conclusion from this line of reasoning.
The Hamiltonian reads as follows;
\begin{eqnarray*}
\mathcal{H}=\frac{1}{2m}\qty(\vb{p}-\frac{q}{c}\vb{A}(\vb{r},t))^2=\frac{1}{2m}\qty(\frac{\hbar}{i}\grad-\frac{q}{c}\vb{A}(\vb{r},t))^2
\end{eqnarray*}
Since the particle is outside the solenoid (if we assume a one-dimensional flux line, there is no longer an interior), we set;
\begin{eqnarray*}
\vb{A}(\vb{r},t)=\vb{A}_{\text{out}}(\vb{r})=\frac{\Phi}{2\pi r^2}(-y,x,0)
\end{eqnarray*}
The wave function $\psi$ depends only on the azimuthal angle $\phi$ (where the polar angle $\theta=\frac{\pi}{2}$ and the radial distance $r=b$ remain fixed) so that the Schrödinger equation reads; 
\begin{eqnarray*}
\frac{1}{2m}\qty[-\frac{\hbar^2}{b^2}\dv[2]{}{\phi}+\qty(\frac{q\Phi}{2\pi b c})^2+i\frac{\hbar q \Phi}{\pi c b^2}\dv{\phi}]\psi(\phi)=E\psi(\phi)
\end{eqnarray*}
In simplified form the latter equation reads;
\begin{eqnarray}
\label{diff_eq}
\dv[2]{\psi}{\phi}-2i\alpha\dv{\psi}{\phi} -\alpha^{2}\psi+\epsilon\psi=0
\end{eqnarray}
where 
\begin{eqnarray*}
\alpha=\frac{q \Phi}{2\pi \hbar c},\qquad
\epsilon=\frac{2m b^2 E}{\hbar^2}
\end{eqnarray*}
A general solution of Eq.~\ref{diff_eq} is;
\begin{eqnarray*}
\psi(\phi)=Ae^{i(\alpha+\sqrt{\epsilon})\phi}+B e^{i(\alpha-\sqrt{\epsilon})\phi}
\end{eqnarray*}
However, the wave equation must be a continuous function, it must therefore fullfill the continuity condition;
\begin{eqnarray*}
\psi(\phi+2\pi)=\psi(\phi), \qquad \forall \phi \in \mathbb{R}
\end{eqnarray*}
This condition is only fullfilled if the following equations hold simultaneously;
\begin{eqnarray*}
\alpha+\sqrt{\epsilon}&=&n,\qquad n\in \mathbb{Z}\\
\alpha-\sqrt{\epsilon}&=&m,\qquad m\in \mathbb{Z}
\end{eqnarray*}
From the first equation, we conclude;
\begin{equation*}
\sqrt{\epsilon}=n-\alpha
\end{equation*}
We substitute $\sqrt{\epsilon}$ into the second equation and obtain;
\begin{equation*}
\alpha-(n-\alpha)=m
\end{equation*}
or equivalently,
\begin{equation}
\label{condition}
\alpha=\frac{n+m}{2}
\end{equation}
which means that $\alpha$ is either a half-integer or an integer.
\begin{equation}
\alpha=\frac{n}{2},\qquad n\in \mathbb{Z}
\end{equation}
 In explicit terms we have the following statement;
\begin{equation}
\label{flux_quantization}
q\cdot \phi=n\cdot\frac{hc}{2},\qquad n\in \mathbb{Z}\quad (h=2\pi \hbar)
\end{equation}
Equation \ref{flux_quantization} is interesting when we read it as a general statement. The magnetic flux and the electric charge are functionally independent in the setting above, so the equation should actually be universally valid; meaning that magnetic flux $\Phi$ and the electric charge $q$ are both quantized quantities. If we denote $e$ as the elementary charge, then $\Phi_{0}=\frac{hc}{2e}$ is the flux quantum known in the context of superconductivity.\\
We also note that the quantization condition of Eq.~\ref{flux_quantization} should hold in any inertial frame, i.e. it should be Lorentz-invariant. From this, and from the theoretically and experimentally well-supported evidence that electric charge is relativistically invariant, one concludes that magnetic flux must be a Lorentz pseudoscalar, \ie relativistically invariant under continuous Lorentz transformations but sign flipping under a space inversion. In the following section, we will show that this is the case.
\section{\label{}Magnetix flux as a Lorentz Pseudoscalar}
As an introduction, we consider the magnetic flux $\Phi$ of a homogeneous, time-independent magnetic field $\vb{B}$ through an infinitesimal surface element $\dd A$ which, for simplicity's sake, should lie entirely in the x-y plane. If we apply a Lorentz boost in an arbitrary direction, the surface element $\dd A$ is contracted by the $\gamma-$factor $\gamma=1/\sqrt{1-{v_{\parallel}^2/c^2}}$ where $v_{\parallel}$ is the projection of the velocity onto the surface element, \ie
\begin{equation*}
{\dd A}^{\prime}=\frac{\dd A}{\gamma}
\end{equation*}
on the other hand, the field component $\vb{B_{\perp}}\equiv B_{3}$ perpendicular to the boost direction transforms as follows;
\begin{equation*}
{B}_{\perp}^{\prime}=\gamma \cdot {B_{\perp}}
\end{equation*}
The magnetic flux $\Phi={B}_{\perp}\cdot \dd{A}={B}_{\perp}^{\prime}\cdot {\dd A}^{\prime}$ remains therefore invariant under a continuous Lorentz transformation. This illustrates the analogous behavior of magnetic flux $\Phi$ and electric charge $q$. Both quantities are invariant under continuous Lorentz transformations.\\
There is a simple way to see why the magnetic flux density $\vb{B}$ - defined as the number of field lines per unit area - increases by a factor of  $\gamma$ under the effect of a Lorentz boost in perpendicular direction: The boost and the resulting length contraction increases the density of field lines (Since field lines live in 3d, being "attached" to space). Therefore, the magnetic flux density $\vb{B}$  behaves in the same way as electric charge density $\rho=\dd q/\dd[3]{x}$.\\ However, the magnetic flux—defined as the total number of field lines passing through the surface element—remains unchanged.
\\
\begin{figure}[h!]
\centering
{\includegraphics[width=0.8\linewidth]
{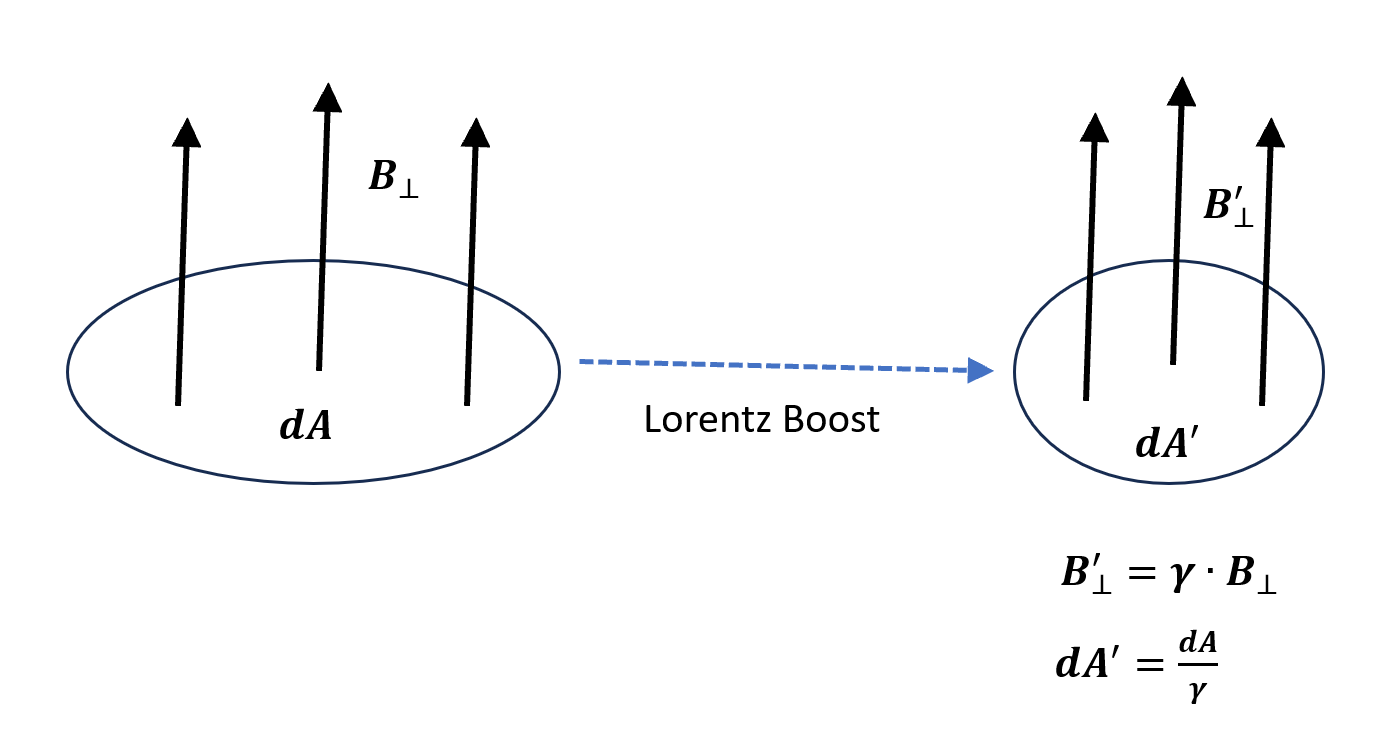}}
\caption{Under a Lorentz boost, the magnetic flux density $\vb{B_{\perp}}$ perpendicular to the boost direction increases by the $\gamma-$factor, but the magnetic flux remains invariant.}
\label{fig2}
\end{figure}
Since magnetic field lines are represented by closed loops (because of $\div \vb{B}=0$), it is intuitively clear that the total magnetic flux through any infinitely extended plane must vanish. Formally, this can be demonstrated by assuming that the fields are vanishing in spatial directions (\ie outside the light cone). First, we express the magnetic flux (via delta functions) as a 4d-Integral in Minkowski space $\mathbb{M}_{4}$; \\
\\
\begin{eqnarray*}
\Phi&=&\frac{1}{c}
\iint\limits_{x_{0}, \newline x_{3}=0}B_{3}(x)\,\dd x_{1}\dd x_{2}=\frac{1}{c}\int\limits_{\mathbb{R}^4}\dd[4]{x}(\grad\cross\vb{A})_{3}\,\,\delta(x_{0})\,\delta(x_{3})\\
\end{eqnarray*}
Next, we write the delta functions as derivatives of Heaviside's unit step functions $\Theta(\cdot)$ defined in the usual way as;
\begin{eqnarray*}
\Theta(x)= \begin{cases}
+1 & \text{if}\,\,x>0\\
\frac{1}{2} & \text{if}\,\,x=0 \\
0 & \text{otherwise}
\end{cases}
\end{eqnarray*}
The magnetic flux $\Phi$ can then be expressed by using the dual field tensor $\tilde{F}^{\mu \nu}$;
\begin{eqnarray}
\label{formula}
\Phi=\frac{1}{c}\int\limits_{\mathbb{R}^4}\dd[4]{x}\tilde{F}^{\mu \nu}\,\partial_\mu\,\Theta(x_{0})\,\partial_\nu\,\Theta(x_{3})
\end{eqnarray}
Note that the following equation holds;
\begin{eqnarray*}
\partial_{\nu}\qty[\tilde{F}^{\mu \nu}\,\partial_\mu\,\Theta(x_{0})\,\Theta(x_{3})]&=&\underbrace{\partial_{\nu}\tilde{F}^{\mu \nu}}_{=0}\qty[\partial_{\mu}\Theta(x_{0})\Theta(x_{3})]\\&+&\tilde{F}^{\mu \nu}[\underbrace{\partial_{\nu \mu}\Theta(x_{0})}_{=0\, \text{for}\, \mu \neq \nu}\Theta(x_{3})+\partial_\mu\,\Theta(x_{0})\,\partial_\nu\,\Theta(x_{3})]\\
&=&\tilde{F}^{\mu \nu}\,\partial_\mu\,\Theta(x_{0})\,\partial_\nu\,\Theta(x_{3})
\end{eqnarray*}
By applying the divergence theorem (Gauss's theorem) we obtain;
\begin{eqnarray*}
\Phi=\frac{1}{c}\int\limits_{\Sigma=\mathbb{R}^4}\dd[4]{x}\partial_{\nu}\qty[\tilde{F}^{\mu \nu}\,\partial_\mu\,\Theta(x_{0})\,\Theta(x_{3})]=\frac{1}{c}\oint_{\partial \Sigma} \tilde{F}^{\mu \nu}\,\partial_\mu\,\Theta(x_{0})\,\Theta(x_{3})\cdot \hat{n} \,\dd[3]S=0
\end{eqnarray*}
The term before the last equality vanishes because - by assumption - the (dual) field tensor vanishes in spatial directions and because of;
\begin{eqnarray*}
\lim_{x_{0}\rightarrow \infty} \partial_\mu\,\Theta(x_{0})=0\qquad \mu \in \{0,1,2,3\}
\end{eqnarray*}
Similarly, it can be shown that the total flux is invariant under continuous Lorentz transformations ${x}^{\prime} = \Lambda x + a$. The transformed flux reads;
\begin{eqnarray*}
{\Phi}^{\prime}=\frac{1}{c}\int\limits_{\mathbb{R}^4}\dd[4]{{x}^{\prime}}{\tilde{F}}^{\prime\mu \nu}\,{\partial}_\mu^{\prime}\,\Theta({x}_{0}^{\prime})\,\partial_{\nu}^{\prime}\,\Theta({x}_{3}^{\prime})
=\frac{1}{c}\int\limits_{\mathbb{R}^4}\dd[4]{{x}}{\tilde{F}}^{\mu \nu}\,{\partial}_\mu\,\Theta({x}^{\prime}_{0})\,{\partial}_\nu\,\Theta({x}^{\prime}_{3})
\end{eqnarray*}
Where we used the fact that the 4-dimensional volume element $\dd[4]{{x}}=\dd x_{0}\dd x_{1}\dd x_{2}\dd x_{3}$ is Lorentz invariant in Minkowski space $\mathbb{M}_4$.
Therefore,
\begin{eqnarray*}
{\Phi}-{\Phi}^{\prime}=\frac{1}{c}\int\limits_{\mathbb{R}^4}\dd[4]{{x}}{\tilde{F}}^{\mu \nu}\,\Big[{\partial}_\mu\,\Theta({x}_{0})\,{\partial}_\nu\,\Theta({x}_{3})-{\partial}_\mu\,\Theta({x}_{0}^{\prime})\,{\partial}_\nu\,\Theta({x}_{3}^{\prime})\Big]
\end{eqnarray*}
Once again, note that;
\begin{eqnarray*}
\partial_{\nu}\qty[\tilde{F}^{\mu \nu}\,\partial_\mu\,\Theta(x_{0}^{\prime})\,\Theta(x_{3}^{\prime})]&=&\underbrace{\partial_{\nu}\tilde{F}^{\mu \nu}}_{=0}\qty[\partial_{\mu}\Theta(x_{0}^{\prime})\Theta(x_{3}^{\prime})]\\&+&\underbrace{\tilde{F}^{\mu \nu}{\partial_{\nu \mu}\Theta(x_{0}^{\prime})}\Theta(x_{3}^{\prime})}_{=0}+\tilde{F}^{\mu \nu}\partial_\mu\,\Theta(x_{0}^{\prime})\,\partial_\nu\,\Theta(x_{3}^{\prime})\\
&=&\tilde{F}^{\mu \nu}\,\partial_\mu\,\Theta(x_{0}^{\prime})\,\partial_\nu\,\Theta(x_{3}^{\prime})
\end{eqnarray*}
The second term vanishes because of;
\begin{eqnarray*}
\tilde{F}^{\mu \nu}&=&-\tilde{F}^{\nu \mu}\\
\partial_{\nu \mu}\Theta(x_{0}^{\prime})&=&\partial_{\mu \nu}\Theta(x_{0}^{\prime})
\end{eqnarray*}
Once again, applying a partial integration;
\begin{eqnarray*}
\Phi^{\prime}=\frac{1}{c}\int\limits_{\Sigma=\mathbb{R}^4}\dd[4]{x}\partial_{\nu}\qty[\tilde{F}^{\mu \nu}\,\partial_\mu\,\Theta(x_{0}^{\prime})\,\Theta(x_{3}^{\prime})]=\frac{1}{c}\oint_{\partial \Sigma} \tilde{F}^{\mu \nu}\,\partial_\mu\,\Theta(x_{0}^{\prime})\,\Theta(x_{3}^{\prime})\cdot \hat{n} \,\dd[3]S=0
\end{eqnarray*}
where - once again - the surface integral vanishes because - by assumption - the fields vanish in spatial directions. In time direction $x_{0}$ we use;
\begin{eqnarray*}
\lim_{x_{0}\rightarrow \infty} \partial_\mu\,\Theta(x_{0}^{\prime})=0\qquad \mu \in \{0,1,2,3\}
\end{eqnarray*}
Note that;
\begin{eqnarray*}
\lim_{x_{0}\rightarrow \infty} x_{0}^{\prime}=\infty
\end{eqnarray*}
The total magnetic flux is therefore a Lorentz pseudoscalar\footnote{The total magnetic flux is invariant under a continuous Lorentz transformation but flips its mathematical sign under a parity transformation (space inversion: $x\rightarrow -x$ or time inversion: $t\rightarrow -t$)}.
\section{Conclusion}
In this paper, we have derived the quantization condition $q\cdot \phi=n\cdot\frac{hc}{2}$ ($n$ an integer) and argued that there is no reason to regard this equation as not universally valid. One consequence would be that both electric charge and magnetic flux are quantized quantities. Furthermore, we have shown that the magnetic flux behaves like a pseudoscalar under a Lorentz transformation. Thus, the total magnetic flux essentially shares (except for parity invariance) the same properties as the electric charge. As a result, the aforementioned quantization condition is Lorentz invariant.
\newpage
\bibliography{magnetic_flux}
\bibliographystyle{plain}
\end{document}